# APPLICATION OF PARTICLE SWARM OPTIMIZATION FOR ENHANCED CYCLIC STEAM STIMULATION IN A OFFSHORE HEAVY OIL RESERVOIR

Xiaolin Wang[1] and Xun Qiu[2]

[1]CNOOC International Ltd., Beijing, P.R. China
wangxl11@cnooc.com.cn
[2]Shenzhen Branch of CNOOC Ltd., Shenzhen, P.R. China
qiuxun@cnooc.com.cn

## ABSTRACT

*Three different variations of PSO algorithms, i.e. Canonical, Gaussian Bare-bone and Lévy Bare-bone PSO, are tested to optimize the ultimate oil recovery of a large heavy oil reservoir. The performance of these algorithms was compared in terms of convergence behaviour and the final optimization results. It is found that, in general, all three types of PSO methods are able to improve the objective function. The best objective function is found by using the Canonical PSO, while the other two methods give similar results. The Gaussian Bare-bone PSO may picks positions that are far away from the optimal solution. The Lévy Bare-bone PSO has similar convergence behaviour as the Canonical PSO. For the specific optimization problem investigated in this study, it is found that the temperature of the injection steam, $CO_2$ composition in the injection gas, and the gas injection rates have bigger impact on the objective function, while steam injection rate and the liquid production rate have less impact on the objective function.*

## 1. INTRODUCTION

Particle Swarm Optimization (PSO) algorithm is co-operative, population-based global search swarm intelligence mataheuristics developed by James Kennedy and Russell C. Eberhart (1995). PSO algorithm has been successfully used as a high efficient optimizer in numerous area. More specifically in petroleum engineering field, PSO have been utilized to perform assisted history matching, optimization in several of recover processes (e.g. Mohamed, et. al., 2010; Onwunalu and Durlofsky, 2011; Zhang, et. al. 2011; Wang, et. al., 2012).

In a Canonical PSO, the movement of each particle is determined by the weighted sum of three components: particle's previous movement direction (inertia), the distance between current position and the local best position (cognition influence) and the distance between current position and the global best position (social influence). Many researchers have proposed modified PSO algorithms to improve the efficiently and robustness of the PSO algorithm. For example, Kennedy and Mendes (2004) suggested a fully informed particle swarm where each particle's movement is affected by all its neighbors instead of just the local optima of itself and the global optima. It has been found that the fully informed PSO perform better in some cases, but the performance depends on the population topology. Because Genetic Algorithm (GA) is similar to PSO in the sense that they are both population based optimization approaches, it is nature to compare the performances of these two methods. Based on Hassan et al. (2004)'s study, PSO requires less computational effort than GA to arrive at the same quality solution. Robinson et al. (2002) found that GA performs better at the beginning of the optimization process while PSO improves faster latter, and thus they proposed a hybrid optimization method which can switch between PSO and GA.

In Canonical PSO and many of its variations, controlling parameters are involved in the calculation process such as inertia factor, social and cognition influence factors. Although many research has suggest constant values for these factors (Eberhart and Shi, 2000, 2001; Clerc and Kennedy, 2002). Proper values of these factor are normally problem specific, thus it is very hard to find the best controlling parameters for these algorithms in real world optimization problems. Bare-bones PSO is a method that does not have any control parameter involved, thus it is easier to be used in real world application. In bare-bones PSO the move of particles are calculated using a probability distribution. Kennedy (2003) first proposed a bare-bones PSO using Gaussian distribution. Latter, Rich and Blackwell (2006) suggested using Lévy distribution, which have a fatter tails compared to Gaussian distribution, to improve the performance of bare-bone PSO. A more detailed review of the PSO algorithm can be found at Poli et. al. (2007).

In literature, many researchers have used various test functions to test PSO algorithms. For example, Yang (2010) provided a very useful list of the commonly used mathematical test functions. However, we believed testing the performance of different algorithm on real world practical problems is also very important and meaningful. In this study, we tested the application of Canonical PSO and bare-bone PSO using a large scale oil field optimization problem. The objective of the study is to maximize the thermal recovery from an off shore heavy oil reservoir. The process we are modeling involves complex chemical and petrophysical processes. The fitness of each particle in PSO is calculated using commercial reservoir simulation software (CMG STARS 2010) to solve the multiple phase flow in porous media.

In this paper, the basic mathematics of the PSO methods is review first. Then the background of the oil field and the objective of the optimization are introduced. Then we summarized the main result of the optimization test obtained from different PSO methods. Finally, some conclusions are draw.

## 2. PARTICLE SWARM OPTIMIZATION ALGORITHMS

### 2.1. Canonical PSO

In order to solve an optimization problem with $D$ decision variables, a swarm with population size $N$ is used. Each particle in this swarm has two vectors: location vector $\boldsymbol{X}=(x^1, x^2, \ldots, x^j, \ldots x^D)$ and velocity vector $\boldsymbol{V}=(v^1, v^2, \ldots, v^j, \ldots v^D)$. The Canonical PSO update the location of a particle $i$ at step $t$+1 using the following two equations:

$$v_i^{j,t+1} = wv_i^{j,t} + c_1 r_1 \left(p_{i,local}^{j,t} - x_i^{j,t}\right) + c_2 r_2 \left(p_{global}^{j,t} - x_i^{j,t}\right) \quad (1)$$

$$x_i^{j,t+1} = x_i^{j,t} + v_i^{j,t+1} \quad (2)$$

Where $i$ is the index of particle; $j$ is the index of solution dimension; $t$ is step(or generation) index; $w$ is inertia factor, $w$=0.7298; $c_1$ is cognition influence factor and $c_2$ is social influence factor, they are both constants, $c_1=c_2$=1.49618; $r_1$ and $r_2$ are two independent random numbers uniformly distributed in the range of [0, 1]; $p_{i,local}$ is the best location found by the $i$th particle so far and $p_{global}$ is the best location found by all particle in the swarm so far.

At initial stage, i.e. $t$=0, the location and velocity vectors are initialized with random values, and then equation 1 and 2 are calculated sequentially and recursively until a preset stop criterion is met.

## 2.2. Gaussian Bare-bone PSO

In Gaussian Bare-bone PSO, the locations of particles are updated using Gaussian distribution.

$$x_i^{j,t+1} = Normal(\mu_i^{j,t}, \sigma_i^{j,t}) \tag{3}$$

Where $\mu_i^{j,t}$ is the mean and $\sigma_i^{j,t}$ is the standard deviation of Gaussian distribution, and they can be calculated using the following equations:

$$\mu_i^{j,t} = \frac{p_{i,local}^{j,t} + p_{global}^{j,t}}{2} \tag{4}$$

$$\sigma_i^{j,t} = \left| p_{i,local}^{j,t} - p_{global}^{j,t} \right| \tag{5}$$

As can be seen in equations 3-5, no control parameters are used in Gaussian Bare-bone PSO, at the same time it can imitate Canonical PSO on many problems. However, in some other cases, Gaussian Bare-bone PSO is not as effective (Richer and Blakwell, 2006).

## 2.3. Lévy Bare-bone PSO

Richer and Blakwell (2006) proposed a new PSO algorithm, which used Lévy distribution to replace the uniform distribution in the Gaussian PSO.

The Lévy distribution, named after Paul Pierre Lévy, is one of the few distributions that are stable and that have probability density functions that are analytically expressible.

Two parameters control the shape of Lévy distribution: scale parameter ($\gamma$) and index of stability or characteristic exponent (σ). In this study, $\sigma = 1.5$ is used. Scale parameter is calculated as:

$$\gamma = 0.568 \left| p_{i,local}^{j,t} - p_{global}^{j,t} \right|^{1.5} \tag{6}$$

Then Lévy random numbers are calculated using a method proposed by Chambers etc. (1976).

# 3. Case Study

## 3.1 Problem Description

The objective of this study is to optimize (maximize) the thermal oil recovery of Cyclic Steam Stimulation (CSS) process in an oil field in Bohai Bay, China. The target reservoir is a structural trap heavy oil reservoir. The basic properties of the reservoir are summaries in Table 1. The depth and the location of perforated wells are shown in Figure 1 and 2, respectively.

Table 1: Reservoir properties

| Property | Unit | Value |
|---|---|---|
| Depth | m | 936.0-1206.0 |
| Thickness | m | 7.0-10.6 |
| Average horizontal permeability | $\mu m^2$ | 3.97 |
| Average porosity | | 34.3% |
| Average oil saturation | | 59.7% |
| Reservoir temperature | °C | 45.6-55.6 |
| Reservoir pressure | MPa | 10.32 |
| Oil viscosity at reservoir condition | mPa·s | 342-584 |
| Oil density at surface condition | $g/cm^3$ | 0.958-0.967 |

## Cyclic Steam Stimulation (CSS)

Base on the experiences of developing similar reservoirs in this area, the final oil recovery factor will be below 10% if a non-thermal recovery method was used. To maximize the oil recovery, it is proposed that CSS technique should be tested in this reservoir.

There are three stages during a CSS process (sometimes called "huff-and-puff"), as shown in Figure 3. Stage one, high-pressure steam is injected into the formation for several days/ weeks which will reduce the viscosity of oils near the wellbore area. Stage two, when a portion of the reservoir is thoroughly saturated, the steam is turned off and the reservoir "soaks" for several days/weeks. This is followed by stage three, the production phase, when the oil is produced from the same wells to the surface. When production rates decline, another cycle of steam injection begins.

Solvent additive thermal recovery techniques have been carried out in many oil fields worldwide. It has been proven that oil recovery can benefit from the additional solvent injected (Srivastava and Castro, 2011; Norman and Brian, 1998). In this study, Enhanced CSS technique is applied, where $CO_2$ and Nitrogen are co-injected with steam into the reservoir to further reduce the oil viscosity and maximize the oil recovery. In order to obtain the maximum oil recovery for this project, it is important to optimize the operational controlling parameters.

## Controlling Parameters and Objective Function

Many factors can affect the performance of the enhanced CSS progress, such as the rate and temperature of the injection fluid, the duration of the soaking stage, the production rate, and the composition of the injecting fluid.

Five parameters are selected as controlling parameters: temperature of the injecting fluid ($t_{inj}$), stream injection rate of injection wells ($q_{steam}$), additive gas ($CO_2$+N2) injection rate of all injection wells ($q_{gas}$), $CO_2$ composition in the additive gas ($C_{CO2}$), and liquid production rate of production wells ($q_{liquid}$). The ranges of these parameters are listed in Table 2. In this study, we assume that all injection wells inject at the same rate while all the production wells produces at the same condition.

The objective function of this study is set as the oil recovery factor after 5 years of production.

Table 2: Control Parameters

| Parameter | Unit | Range |
|---|---|---|
| $t_{inj}$ | °C | 150-350 |
| $q_{steam}$ | $m^3$ | 150-400 |
| $q_{gas}$ | $m^3$ | 43200-72000 |
| $C_{CO2}$ | | 0.0-0.2 |
| $q_{liquid}$ | $m^3$ | 150-400 |

## 4. RESULTS AND DISCUSSION

In this study all three PSO methods are used to optimize the recovery process of this reservoir. The swarm population size if set to be 10. It should be noted that the simulation job is computationally intensive. Each job takes 1-2 hours to finish on a computer with 12 core CPU. Thus, a maximum of 100 runs (i.e. 10 iterations) were carried out for each methods.

The optimization progresses of the three PSO methods are shown in Figure 4 to Figure 6, respectively. The red diamond symbols demonstrate the optimal solution obtained. It can be seen that all algorithms converge to an area with relative high oil recovery factors, which indicate that the optimization algorithm is effective of finding optimum values within a small amount of simulation runs.

For Canonical PSO, the best run is the 91$^{th}$ run with a recovery factor of 19.64. Compared to base case (Recovery factor is 14.43), the objective function increased by 36.11%. It can been seen in Figure 4 that Canonical PSO converged is very fast, a much higher objective function can be found only after 2-3 iterations. After that, the optimization method mainly searches around the optimal area and improve the objective function. The optimal solution found by Canonical PSO has the highest value among all three PSO methods.

For Gaussian PSO, the best run is found to be Run 77 with a recovery factor of 19.46. The final objective function is a little bit less than that found by the Canonical PSO; however, the steam injection temperature of the optimal run is the lowest among all optimal solutions found by three PSO optimization methods. This operation parameter is more favourable to the field engineers because a lower injecting temperature will be much more cost effective compared to a higher injecting temperature. The optimization process is shown in Figure 5. It is obvious that, the Gaussian PSO method sometime picks parameters that are far away from global optimal cases. This makes the Gaussian PSO less effective in terms of computational cost.

For Lévy Bare-bone PSO, the best run is found at to be number 75 with a recovery factor of 19.31. This result demonstrated that the Lévy Bare-bone PSO is also effective in this large field optimization problem. In addition, it did not require any tuning parameters compared to the Canonical PSO. The convergence behaviour of Lévy Bare-bone PSO is shown in Figure 6. Compared to the Gaussian Bare-bone PSO method, the Lévy Bare-bone PSO has a much stable convergence behaviour, which similar to the Canonical PSO. This is mainly ascribe to the fact that the Lévy distribution has a fat tail kurtosis compared to the normal distribution. This fat tail will can enhance the particles to move out of local sub-optimal positions. The optimal parameter sets found are quite different from that of Canonical PSO method. The optimal steam injection rate and water production rate were much higher than the optimal values found by Canonical PSO. This clearly showed that this optimization problem has multiple optimal. In addition, the impact of steam injection rate and liquid production rate on objective function are not as big as other three parameters.

Table 3: optimization results compared to base case

| PSO Algorithm | Optimal Run ID | Optimal Oil Recovery Factor(%) | Parameters | | | | |
|---|---|---|---|---|---|---|---|
| | | | $t_{inj}$ | $q_{steam}$ | $C_{CO2}$ | $q_{liquid}$ | $q_{gas}$ |
| Base case | 0 | 14.43 | 240 | 210 | 0.12 | 200 | 57600 |
| Canonical | 91 | 19.64 | 350 | 338.986 | 0.2 | 163.747 | 71657.39 |
| Gaussian Bare-bone | 77 | 19.46 | 307.11 | 348.54 | 0.2 | 206.28 | 72000 |
| Lévy Bare-bone | 75 | 19.31 | 349.8946 | 379.6046 | 0.199087 | 400 | 71999.95 |

To verify the above findings, the relationship between the objective function and each parameter are plotted in Figure 7. We only showed the runs checked in Canonical PSO. Figure 7(a) shows that a higher steam temperature will lead to a higher oil recovery factor. This is because the higher the injection steam temperature, the lower the in situ oil viscosity will be, consequently, the mobility of oil phase will be improved. However, it should be noted that, the objective functions are very low when the steam temperature is below 300 °C; once the temperature is above 300°C, there is a jump in the objective function. Thus it is recommended that the steam temperature should be maintained above 300 °C in filed operation.

Figure 7(b) shows a more scatter distribution of points, which indicates that the objective function is not very sensitive to steam injection rate. A similar plot is shown in Figure 7(d). These two plots demonstrated that the oil recovery factor of this field is not very sensitive to steam injection rate and the liquid production rate. A possible reason for this finding may related to the constant soak and production period used in this study, more work need to be done to consider different soak and production time to further investigate the impact of the steam injection and liquid production rates.

Figure 7(C) shows that there is a linear relationship between the $CO_2$ composition in steam and the oil recovery factor. The higher the $CO_2$ percentage, the higher the final oil recovery factor is. This result demonstrated that $CO_2$ is an effective solver to help recover more heavy oil out of this field. The additive $CO_2$ in steam will greatly reduce the oil viscosity and improve the oil recovery. For this case, $CO_2$ additive steam injection is a much favourable thermal recovery method compared to pure steam injection.

Figure 7(e) shows that a higher gas injection rate will result in a high oil recovery factor. This is because at higher injection rates the $CO_2$ injection rate will also be higher, which in turn increase the oil recovery.

## 5. CONCLUSIONS

Three different types of Particle Swarm Optimization Algorithm are tested on a large scale reservoir simulation problem. It is found that all types of PSO algorithm are effective in terms of searching for optimal objective functions. In this study, the highest objective function is found by using the Canonical PSO algorithm, while the Gaussian Bare-bone and Lévy Bare-bone also found similar optimal objective functions. It is found that Gaussian PSO method some time picks parameters far away from optimal values. The Lévy Bare-bone is recommended to be used for similar problem because it provides stable convergence behaviour similar to the Canonical PSO, at the same time, it does not need control parameter tuning.

For this specific reservoir optimization problem, it is found that steam injection temperature should be higher than 300°C to get high recovery factor, and the oil recovery increased almost linearly as the $CO_2$ composition in the injection gas increases. A higher gas injection rate will also yield higher oil recovery. The steam injection rate and the liquid production rate have less impact on the final oil recovery factor. More studies are needed to incorporate more control parameters, e.g. soaking time, into the optimization process to further investigate the combined impact of those operational parameters.

## ACKNOWLEDGEMENTS

The authors would like to thank China National Offshore Oil Corporation for the permission to publish this paper.

## AUTHORS

**Xiaolin Wang** holds B.S. and Master Degrees in Petroleum Engineering from China University of Petroleum, East China. He has 7 years experience in conduction research and development projects for offshore oil and gas reservoir development. He now works as a reservoir engineer in CNOOC International Ltd. His current research interests are in the fields of thermal reservoir simulation, and application of artificial intelligence in reservoir optimization and uncertainty analysis.

**Xun Qiu** holds a B.S. Degree in Applied Mathematics (minor in Petroleum Engineering) from China University of Petroleum, East China. She now works as an intermediate economic engineer in Shenzhen Branch of CNOOC Ltd.

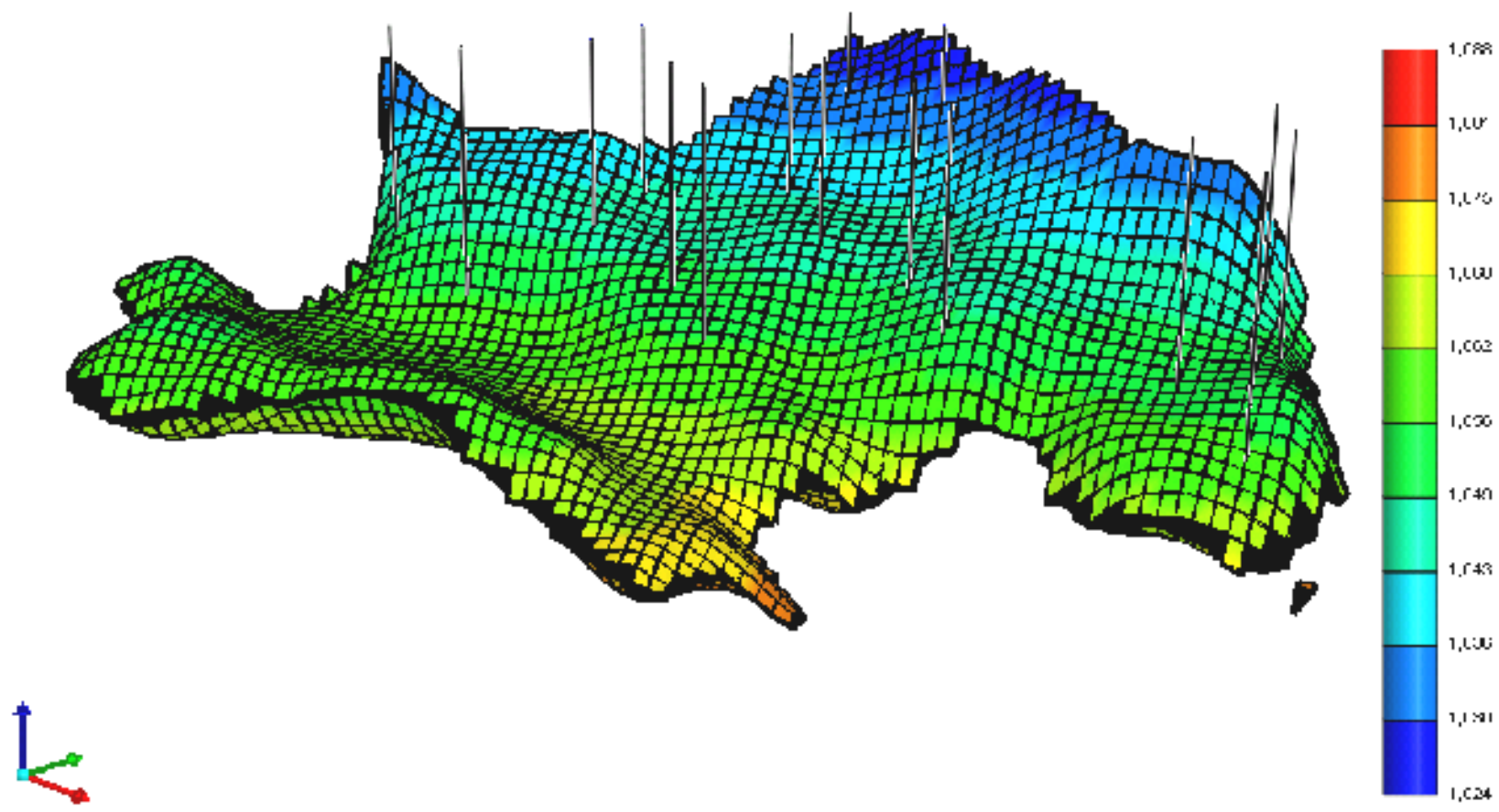

Figure 1. Depth of the reservoir model.

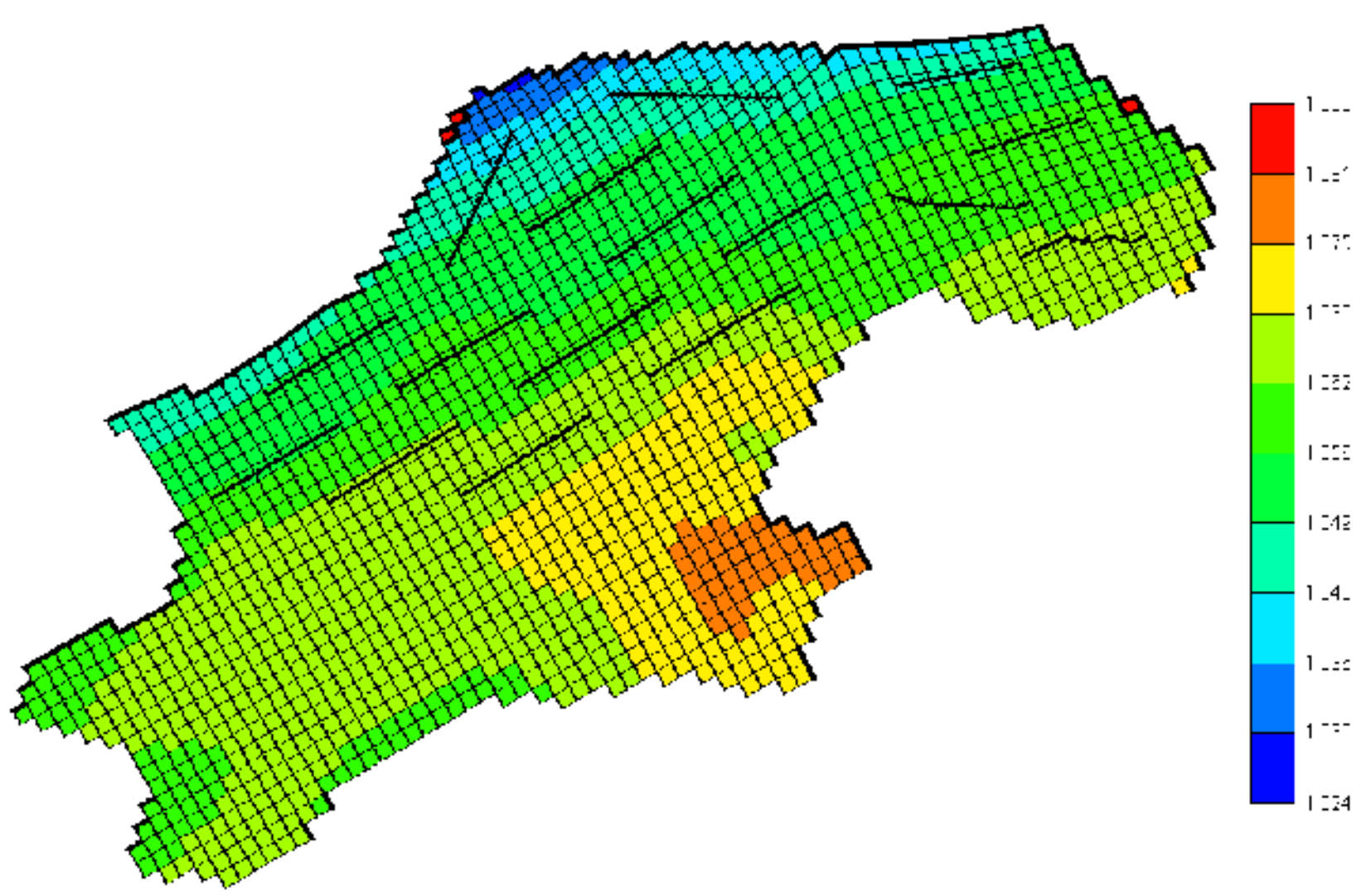

Figure 2. Perforation location of horizontal wells in the reservoir.

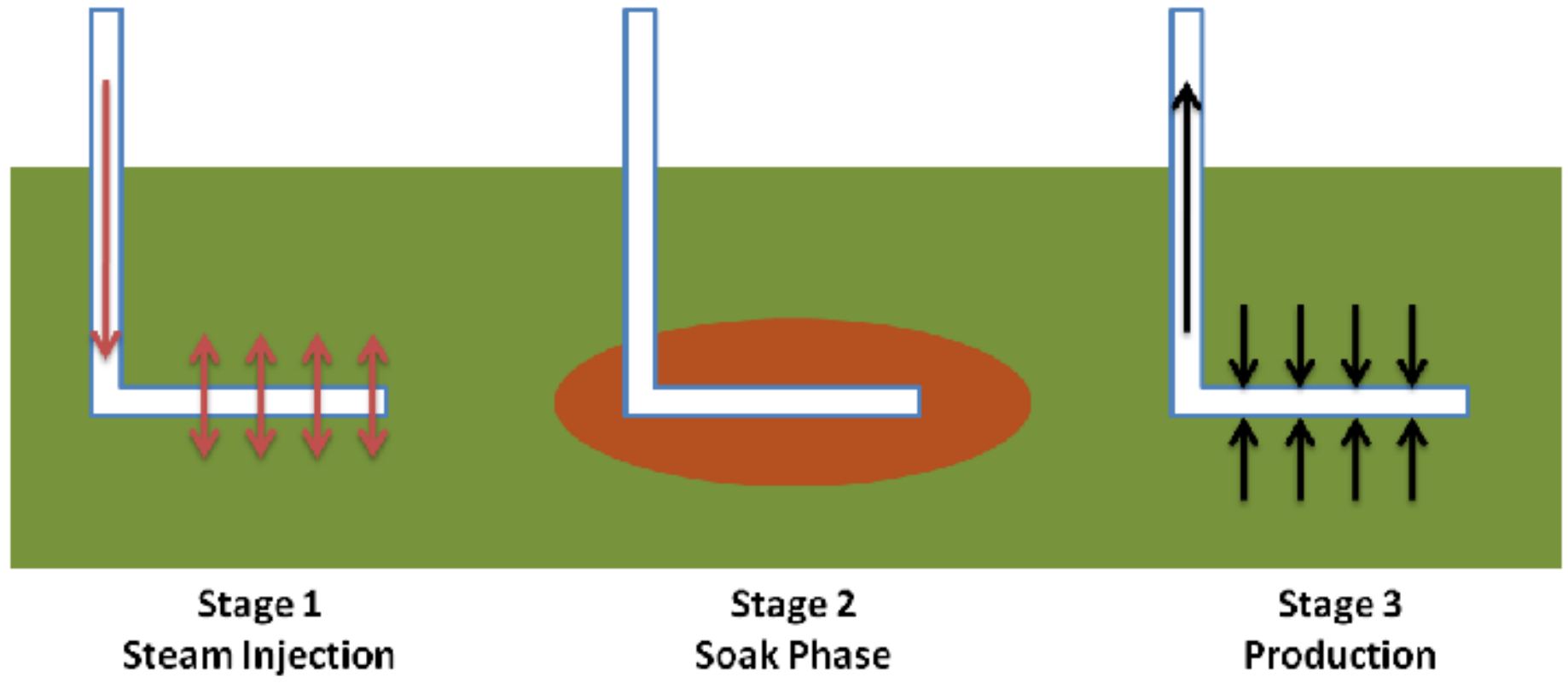

Figure 3. Illustration of Cyclic Steam Stimulation process.

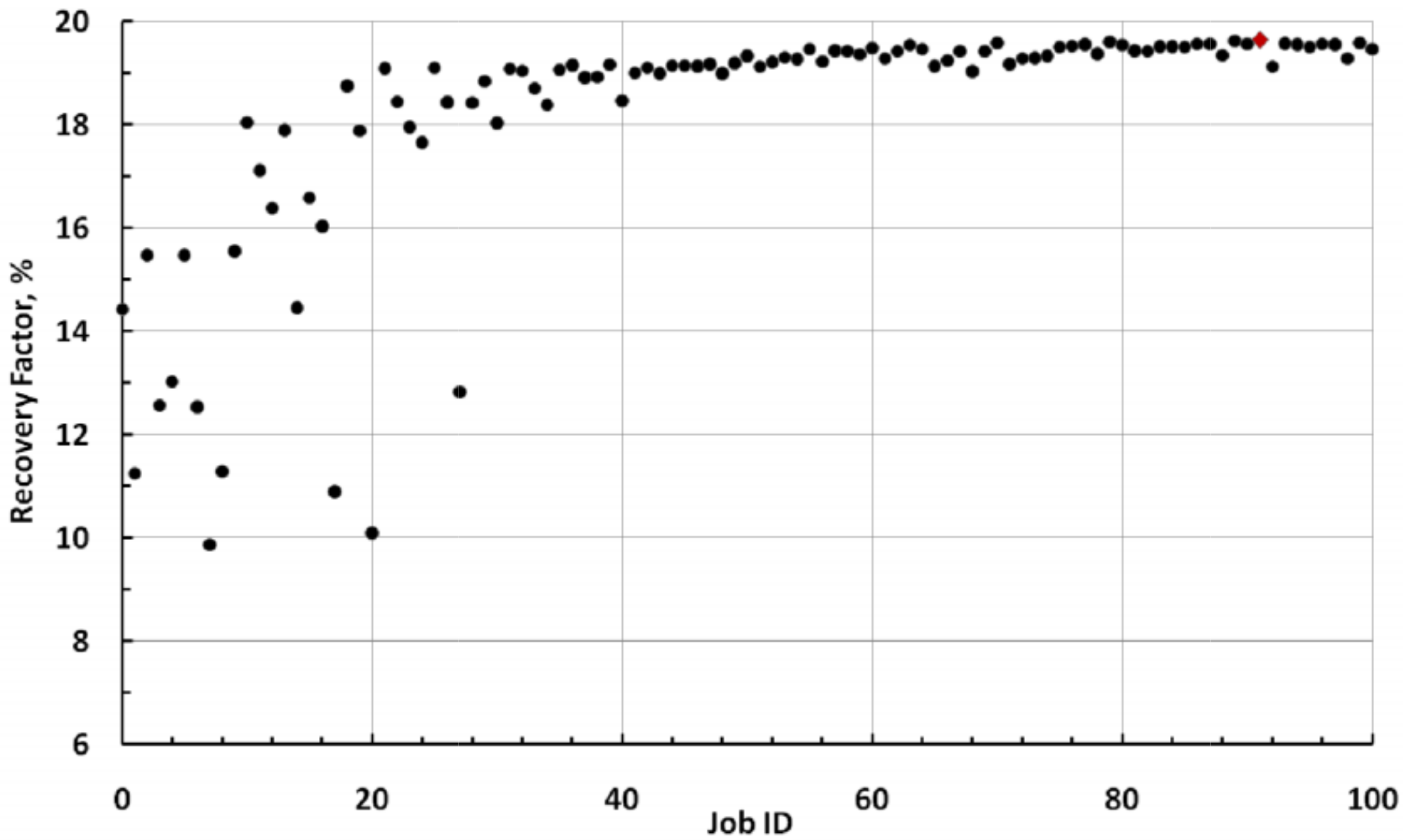

Figure 4. Optimization process using Canonical PSO (optimal job 91 is shown as red diamond)

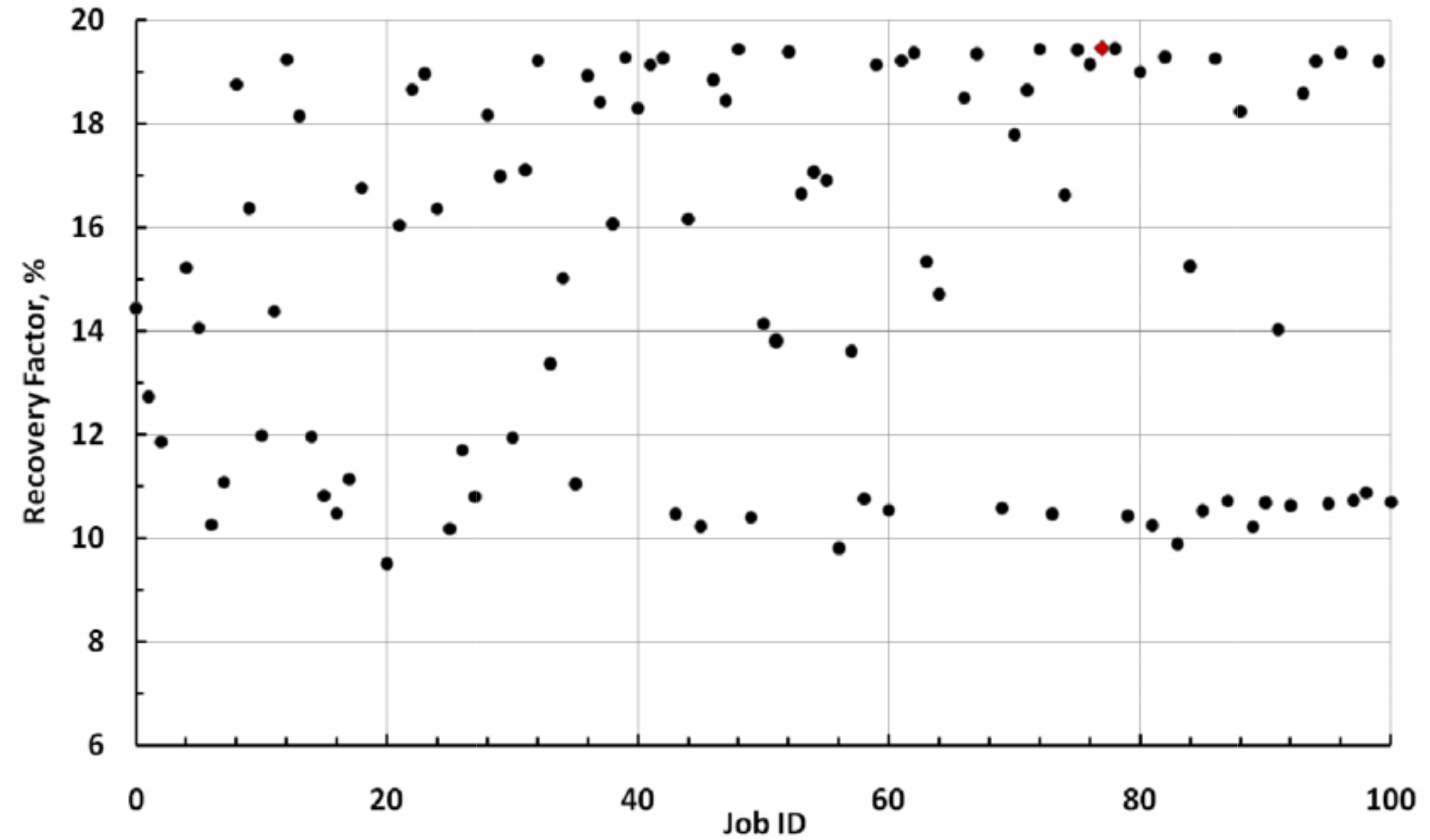

Figure 5. Optimization process using Gaussian Bare-bone PSO (optimal job 77 is shown as red diamond)

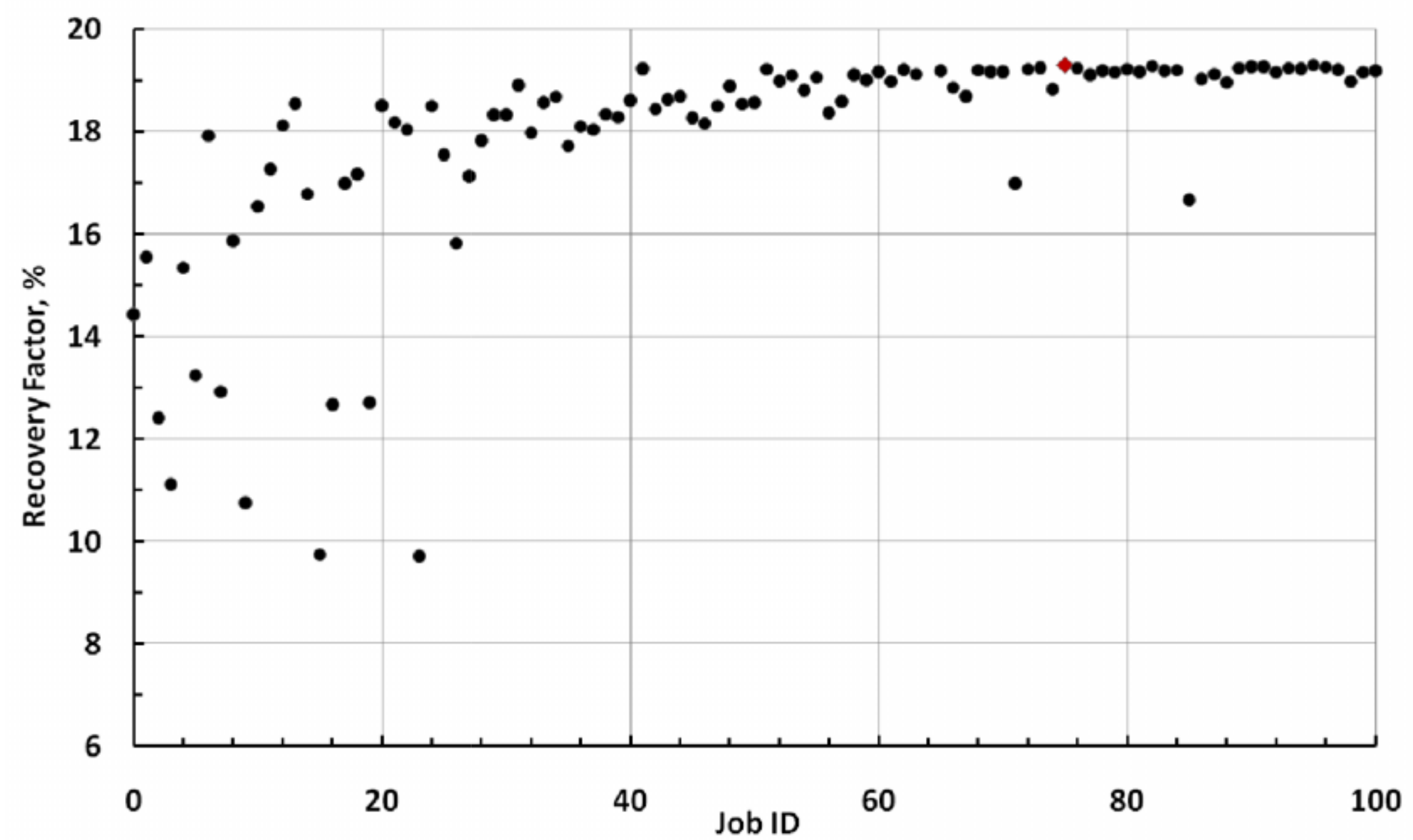

Figure 6. Optimization process using Lévy Bare-bone PSO (optimal job 75 is shown as red diamond)

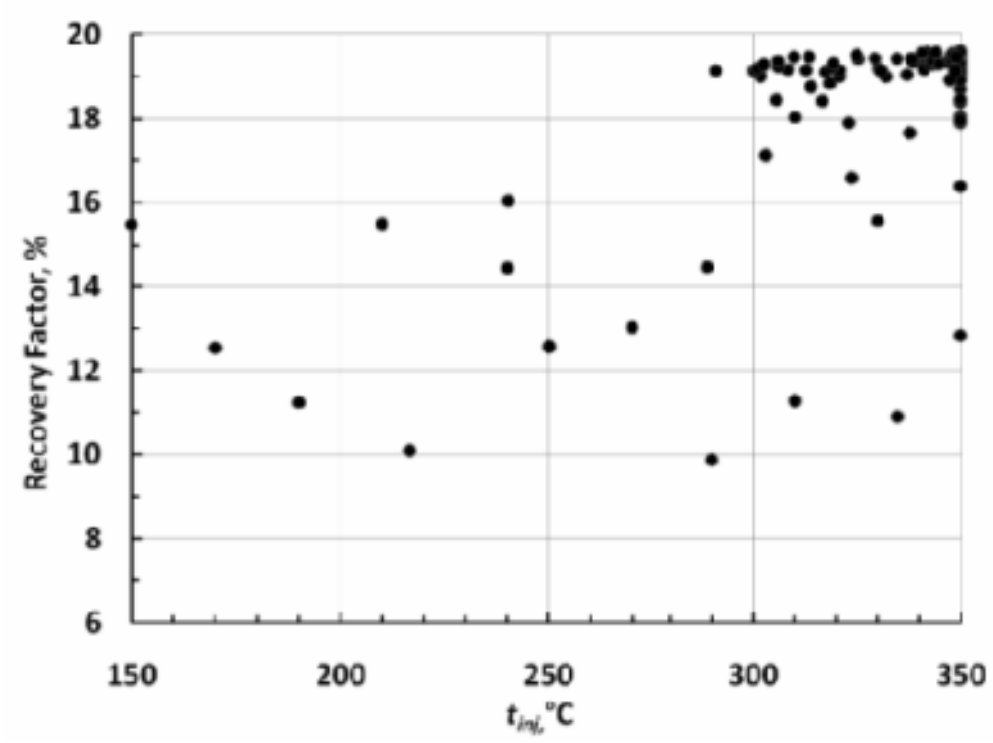

(a) $t_{inj}$

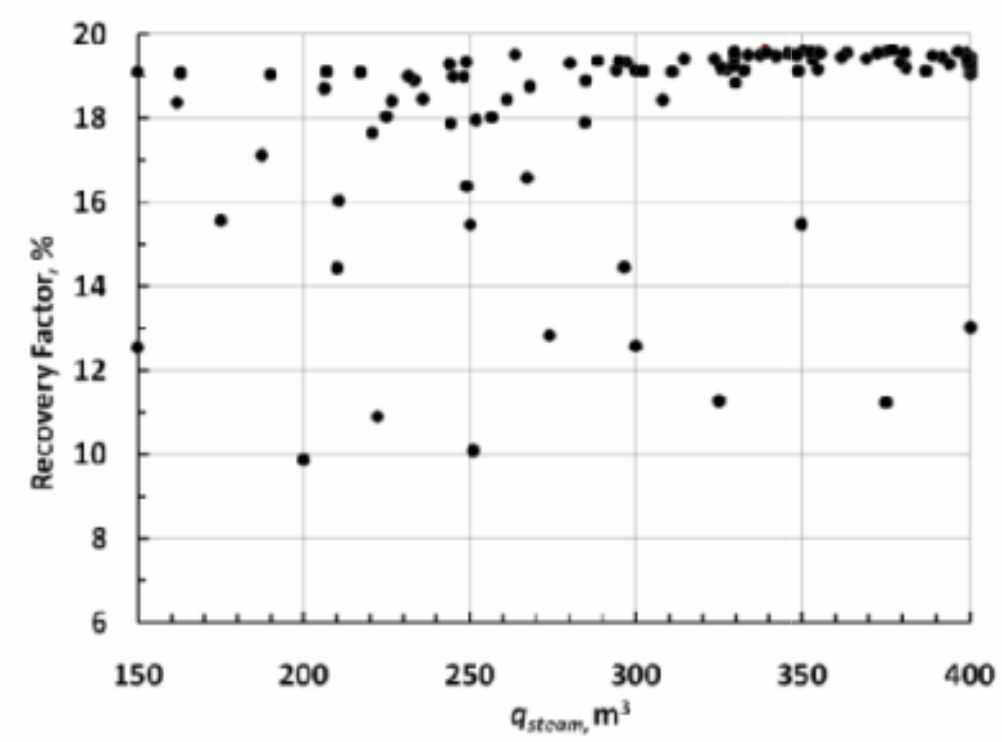

(b) $q_{steam}$

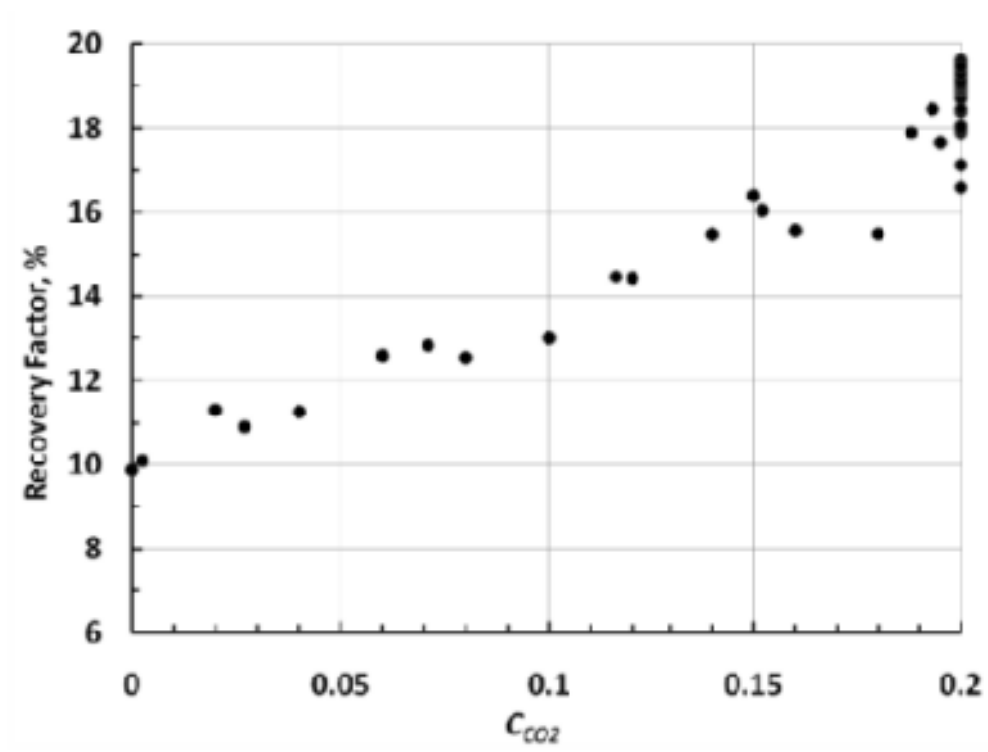

(c) $C_{CO2}$

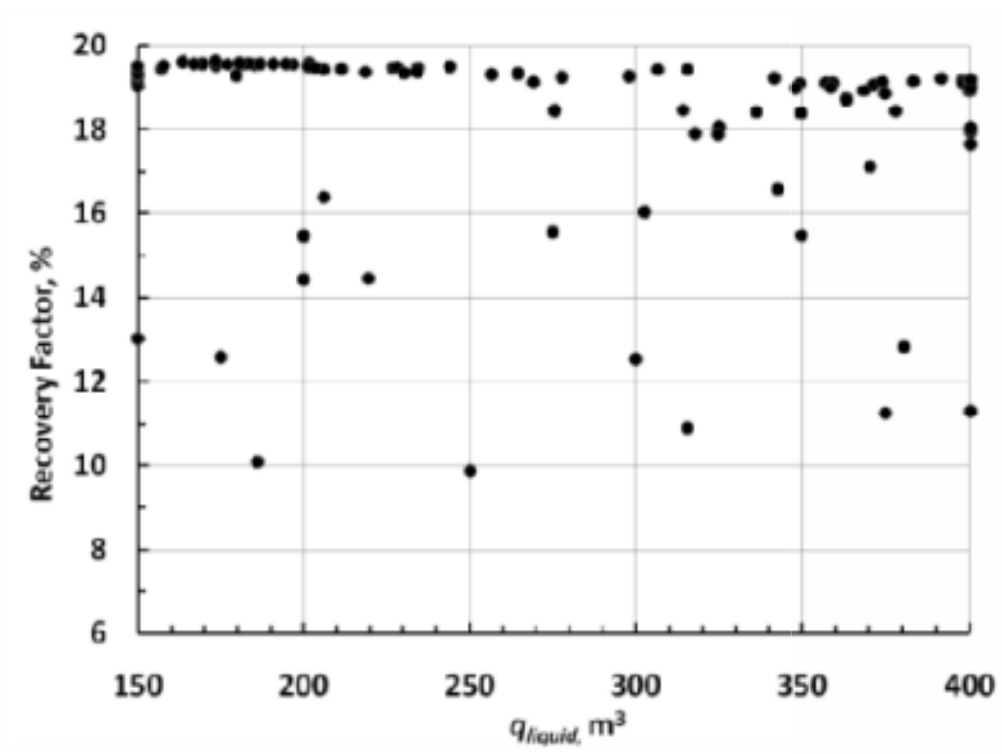

(b) $q_{liquid}$

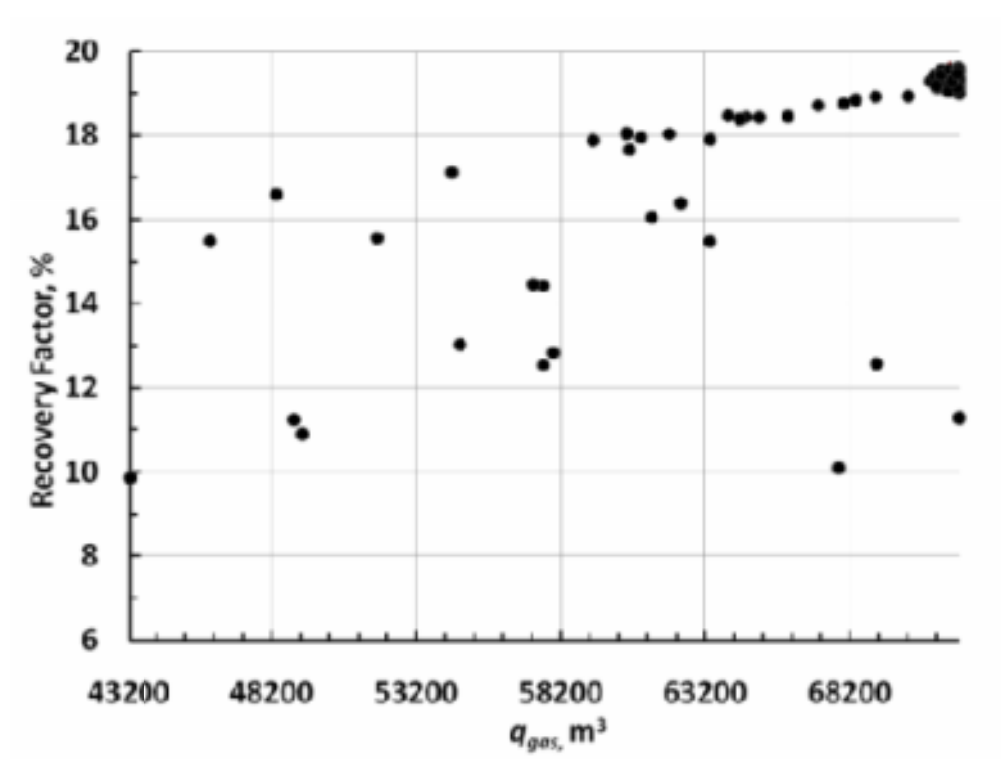

(e) $q_{gas}$

Figure 7. Relationship between the objective function and each parameter (Canonical PSO)